# Toward the End of Stars: Discovering the Galaxy's Coldest Brown Dwarfs

An Astro2010 Decadal Review Science White Paper

prepared by

Adam J. Burgasser* (Massachusetts Institute of Technology)
Josh Bloom (UC Berkeley)
Kelle Cruz (California Institute of Technology)
Michael Cushing (U. Hawaii Institute for Astronomy)
Sandy Leggett (Gemini)
Katharina Lodders (Washington University)
Amanda Mainzer (NASA JPL)
Mark Marley (NASA AMES)
Stanimir Metchev (SUNY Stony Brook)
Subhanjoy Mohanty (Imperial College, UK)
Ben Oppenheimer (American Museum of Natural History)
Andrew West  (Massachusetts Institute of Technology)

*Contact information: (phone) 617 452 5113 (email) ajb@mit.edu

**Brown dwarfs** are the lowest-mass population of stars, distinguished by their inability to sustain core hydrogen fusion and consequent secular evolution to low luminosities and photospheric temperatures ($T_{ph}$).  With masses extending from ~0.075 $M_\odot$ (the hydrogen-burning minimum mass, or HBMM) to below ~0.013 $M_\odot$ (the deuterium-burning minimum mass, DBMM), brown dwarf populations sample the low-mass limits of star formation processes, and are both numerous and ubiquitous in the vicinity of the Sun and throughout the Galaxy at large.  The discovery of brown dwarfs in the 1990s was cited in NAS's 2001 *"Astronomy and Astrophysics in the New Millennium"* report as one of the highlights of that era, and a critical development for future studies of star formation.  The field has advanced significantly since then, with hundreds of brown dwarfs identified and two new spectral classes—the L dwarfs and the T dwarfs—defined to encompass the lowest-temperature examples (Kirkpatrick 2005, *ARA&A* **43**, 195). Studies of brown dwarf populations have provided discriminating constraints on star and planet formation theories, motivated advances in gas chemistry and cloud formation in low-temperature atmospheres, challenged our understanding of magnetic field generation in fully convective interiors, and stimulated work on atomic line-broadening and molecular opacities.  In addition, low-temperature brown dwarfs have atmospheres similar to those of gas giant extrasolar planets, but are typically isolated and therefore more easily studied.  This leads to considerable synergy between the brown dwarf and exoplanet scientific communities, with observations from the former providing ground truth for models and guiding direct detection strategies for the latter.

This White Paper to the National Academy of Sciences Astro2010 Decadal Review Committee highlights cross-disciplinary science opportunities over the next decade with **cold brown dwarfs**, sources defined here as having photospheric temperatures less than ~1000 K.  Key astrophysical science questions relevant to these objects include:

- What are the characteristics of low-temperature atmospheres ranging from gas giant planets in the solar system to low-mass stars?  How do atmospheric processes of condensate cloud formation, gas dynamics, and non-equilibrium chemistry change according to temperature, elemental abundance, mass and age variations?

- What is the contribution of cold brown dwarfs to the Galaxy's thin disk, thick disk and halo populations, and what can we infer about the bulk efficiency and history of brown dwarf formation in the Galaxy?

- How do we observationally determine the physical properties of individual cold brown dwarfs, and how accurate are current atmospheric and evolutionary models that guide both brown dwarf and exoplanet research?

We outline here several investigations feasible over the next decade that could address these science questions, and the infrastructure needs that would support them.

**Understanding Low-Temperature Atmospheres: Constraints from Brown Dwarfs**

The synergy between brown dwarf and exoplanet science is fostered by common problems associated with low-temperature atmospheres. Equilibrium and non-equilibrium chemistry, condensate grain cycles and cloud formation, and feedback mechanisms on atmospheric pressure-temperature profiles (including irradiation) are all issues currently under investigation by observers and theorists in the brown dwarf and exoplanet communities.  It is with brown dwarfs, however, that we can obtain the most robust empirical tests of models for temperatures spanning those of Jupiter (~125 K) to the coldest known brown dwarfs (~575 K; Burningham et al. 2008, *MNRAS* **391**, 320).

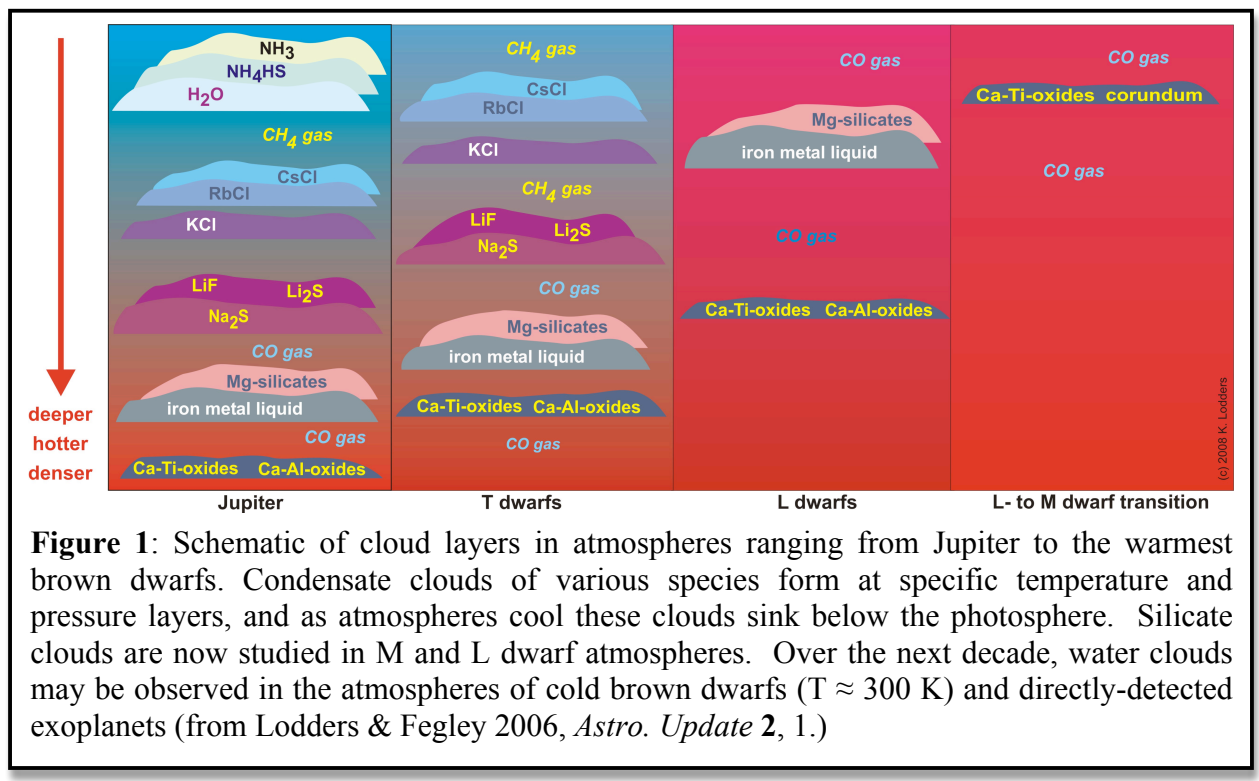

**Figure 1**: Schematic of cloud layers in atmospheres ranging from Jupiter to the warmest brown dwarfs. Condensate clouds of various species form at specific temperature and pressure layers, and as atmospheres cool these clouds sink below the photosphere. Silicate clouds are now studied in M and L dwarf atmospheres. Over the next decade, water clouds may be observed in the atmospheres of cold brown dwarfs (T ≈ 300 K) and directly-detected exoplanets (from Lodders & Fegley 2006, *Astro. Update* **2**, 1.)

An important but poorly understood process in the atmospheres of brown dwarfs and gas giant planets is the formation and evolution of condensate clouds. Chemical equilibrium models predict the formation of mineral condensates in the photospheres of L-type dwarfs, including refractory ceramics (Ca-aluminates, Ca-titanates), silicates, and liquid iron (**Figure 1**; Allard et al. 2001, *ApJ* **556**, 357; Lodders 2002, *ApJ* **577**, 974). These condensates are vertically constrained into cloud layers by diffusion, grain growth, settling and evaporation cycles (Ackerman & Marley 2001, *ApJ* **556**, 872; Helling & Woitke 2006, *A&A* **455**, 325). Condensate cloud formation processes are observationally inferred from cloud opacity effects on spectral energy distributions (Tsuji et al. 1996, *A&A* **380**, L29; Knapp et al. 2004, *ApJ* **127**, 3553), modified elemental abundances and chemistry above the cloud layers (Burrows & Sharp 1999, *ApJ* **512**, 843), and direct detection of silicate grain absorption features (Cushing et al. 2006, *ApJ* **648**, 614).

Yet this conceptual picture of cloud formation remains incomplete. Current models are unable to describe large-scale cloud structure and associated photometric and spectral variability (Gelino et al. 2002 *ApJ*, **577**, 433; Goldman et al. 2008, *A&A* **487**, 277), fail to reproduce the rapid disappearance of clouds at the transition between the L dwarf and T dwarf classes (Dahn et al. 2002, *AJ* **124**, 1152; Liu et al. 2006, *ApJ* **647**, 1393; Burgasser 2007, *ApJ* **659**, 655), and do not address the physical basis underlying variations in cloud properties (Cushing et al. 2008, *ApJ* **678**, 1372). Detailed cloud physics, such as grain size distributions, heterogeneous grain compositions, feedback on pressure-temperature profiles, and the interplay between chemistry and gas transport are only beginning to be addressed (Helling et al. 2008, *A&A* **485**, 547; Saumon & Marley 2008, ApJ **689**, 1327). These investigations will likely require the adaptation of Global Circulation Models (GCM), which are now being employed for exoplanet atmospheric studies (e.g., Showman et al. 2008, *ApJ* **682**, 559). GCM models are also needed to fully understand atmospheric gas dynamics in brown dwarf atmospheres, traced by nonequilibrium abundances of $CO/CH_4$ and $N_2/NH_3$ (Griffith & Yelle 1999, *ApJ* **519**, L85; Saumon et al. 2006, *ApJ* **647**, 552).

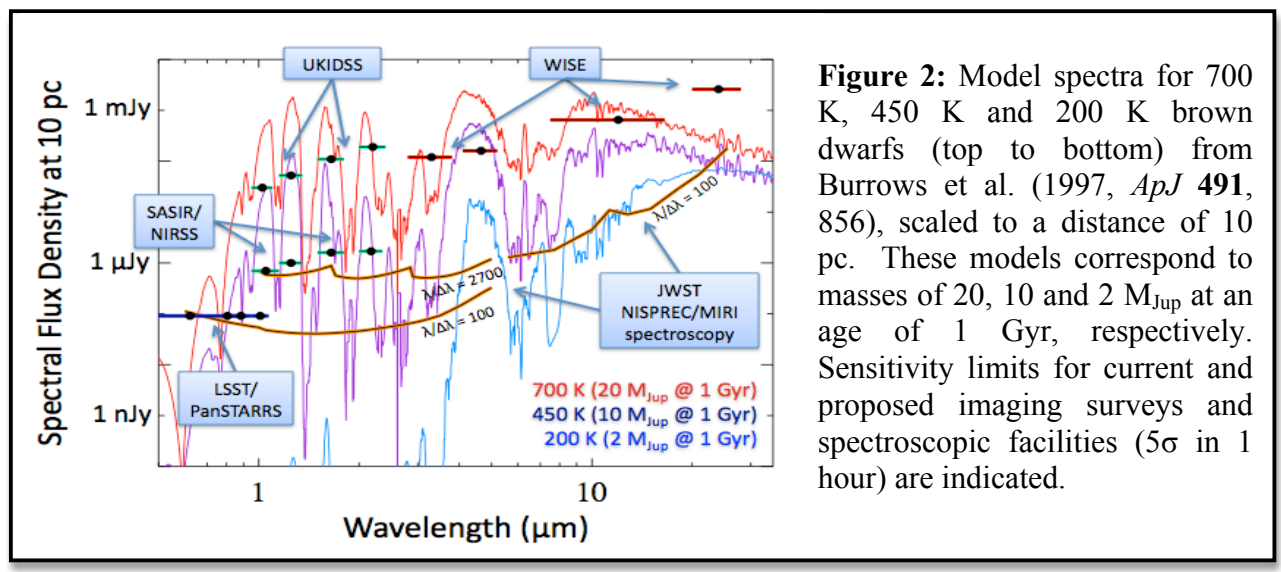

**Figure 2:** Model spectra for 700 K, 450 K and 200 K brown dwarfs (top to bottom) from Burrows et al. (1997, *ApJ* **491**, 856), scaled to a distance of 10 pc. These models correspond to masses of 20, 10 and 2 $M_{Jup}$ at an age of 1 Gyr, respectively. Sensitivity limits for current and proposed imaging surveys and spectroscopic facilities (5σ in 1 hour) are indicated.

Limitations in our current understanding of condensate cloud formation and atmospheric dynamics affect future studies of colder brown dwarfs and directly detected exoplanets. In lower-temperature atmospheres, new condensate species appear—alkali salts (~500-700 K), water (~300 K), ammonia (~200 K) and $NH_4SH$ (~200 K)—each likely to have their own grain formation cycles and influence on spectral energy distributions and pressure-temperature profiles (Lodders & Fegley 2002, *Icarus* **155**, 393). These condensates will be present in the low-temperature atmospheres of extrasolar planets well-separated from their host star, and hence most amenable to direct study. High, reflective clouds increase planetary albedos, modifying the thermal balance of irradiated planets and their detectability (Marley et al. 1999, *ApJ* **513**, 879; Sudarsky et al. 2003, *ApJ* **588**, 1121). With several missions over the next decade aimed at direct studies of exoplanetary atmospheres, accurate models of cloud formation are essential. Again, cold brown dwarfs provide the necessary empirical guides for this work.

To advance studies of low-temperature atmospheres, cold brown dwarfs must first be found. Current near-infrared (NIR) and mid-infrared (MIR) survey programs such as UKIDSS and WISE are expected to provide multi-wavelength detection of $T_{ph} \approx$ 400-500 K brown dwarfs out to 10 pc in the next few years (**Figure 2**). WISE should also make the first single-band, multi-epoch detections of $T_{ph} \approx$ 300 K "water-cloud" brown dwarfs. These discoveries can be verified and cooler brown dwarfs found with panchromatic data provided by deep optical/NIR surveys achieving ~0.1-1 μJy sensitivities, e.g., LSST, Pan-STARRS, JDEM, SASIR, and NIRSS. However, to reach $T_{ph} \approx$ 200 K "ammonia cloud" brown dwarfs, a MIR, wide-field, follow-on survey to WISE is needed, reaching ~10x deeper at 5 μm. Instrumentation sufficient to measure the spectra of these sources will be required to characterize their atmospheres; the MIR is particularly important for studies of $T_{ph}$ < 300 K brown dwarfs. JWST's NIRSPEC and MIRI and SPICA's BLISS instruments will be crucial for this work.

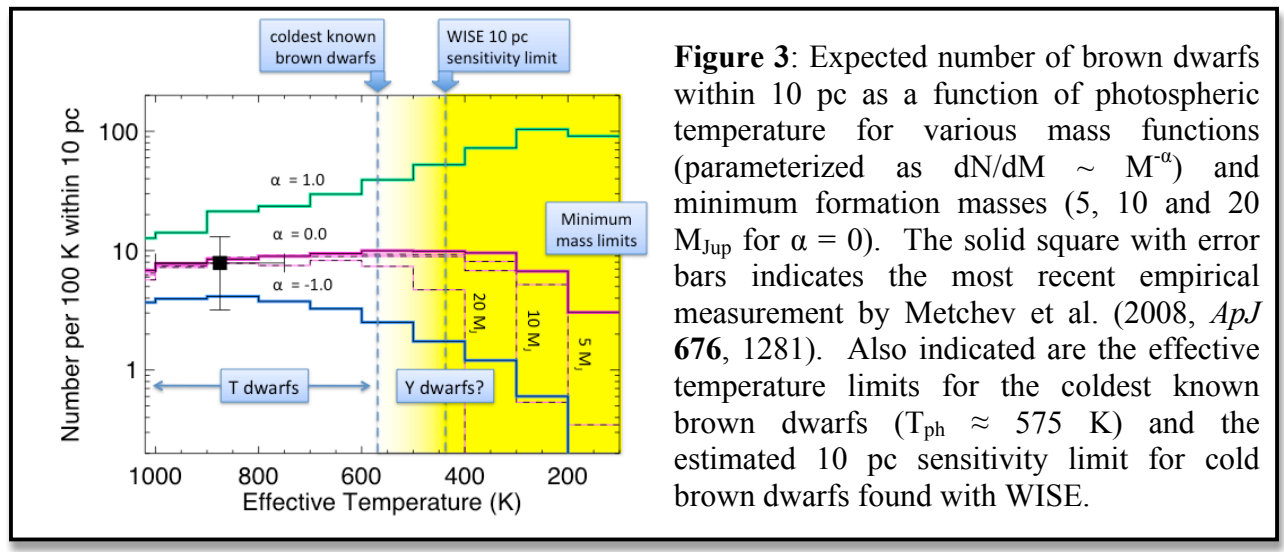

**Figure 3**: Expected number of brown dwarfs within 10 pc as a function of photospheric temperature for various mass functions (parameterized as dN/dM ~ $M^{-\alpha}$) and minimum formation masses (5, 10 and 20 $M_{Jup}$ for $\alpha$ = 0). The solid square with error bars indicates the most recent empirical measurement by Metchev et al. (2008, *ApJ* **676**, 1281). Also indicated are the effective temperature limits for the coldest known brown dwarfs ($T_{ph} \approx$ 575 K) and the estimated 10 pc sensitivity limit for cold brown dwarfs found with WISE.

**Galactic Populations of Brown Dwarfs and Tests of Low Mass Star Formation**

The size of the local cold brown dwarf population constrains the low-mass extremum of the Galactic stellar initial mass function (IMF), the number of stars and brown dwarfs formed in the Galaxy as a function of mass. While reasonable constraints on the shape of the IMF down to the DBMM have been made in the youngest (1-5 Myr) star-forming regions (e.g., Muench et al. 2002, *ApJ* **573**, 366), it is not yet clear that these systems produce the bulk of the Galactic stellar population (Adams et al. 2001, *ApJ* **553**, 744; Lada & Lada 2003, *ARA&A* **41**, 57). A measure of the local brown dwarf IMF is therefore essential for Galactic star formation theories, and hinges on an accurate census of the lowest-temperature sources (Burgasser 2004, *ApJS* **155**, 191; Allen et al. 2005, *ApJ* **625**, 385). The size of the cold brown dwarf population is currently unknown to within 1-2 orders of magnitude (**Figure 3**; Metchev et al. 2008, *ApJ* **676**, 1281) as warm brown dwarf samples provide relatively weak constraints on the form of the IMF. However, population numbers diverge widely for different IMFs below $T_{ph} \approx$ 500 K, so surveys sensitive to such objects out to several tens of pc would greatly improve statistical measures of the IMF. Low-temperature brown dwarfs could also probe the minimum mass of Galactic star formation, which is imprinted as a cutoff in observed number counts at the temperature of the lowest mass object at the age of the Galaxy (Burgasser 2004, *ApJS* **155**, 191). For a minimum formation mass of 10 $M_{Jup}$, simulations predict a decline at $T_{ph} \approx$ 300 K, potentially achievable in the coming decade.

For warmer brown dwarfs, sensitive, large-volume surveys will improve statistics on the local thin disk population and enable studies of older Galactic populations. Currently, $T_{ph} \approx$ 1000 K brown dwarfs can be detected out to ~50-100 pc (e.g., UKIDSS; Pinfield et al. 2008, *MNRAS* **390**, 304), so samples are dominated by the thin disk population. A benchmark goal for the next decade is to extend these samples out to ~500-1000 pc, requiring a 100-fold increase in sensitivity. As the mean age of the disk increases for larger distances above or below the Galactic plane (Loebman et al. 2008, *AIP* **1082**, 238; West et al. 2008, *AJ* **135**, 785), deep surveys of brown dwarfs will more broadly sample age and evolutionary effects in the thin disk. Such surveys will also probe the thick disk and halo brown dwarf populations with greater statistical reliability than current small and largely serendipitous samples (e.g., Burgasser et al.

2007, *ApJ* **657**, 494). Indications of a modified IMF have been found in extremely metal-poor stellar populations in the Galaxy (Komiya et al. 2007, *ApJ* **658**, 367). A measurement of the Galactic halo IMF down to the hydrogen burning limit could test this idea at lower masses.

A NIR survey sensitive to ~1 µJy, such as those planned by JDEM, SASIR and NIRSS, would detect $T_{ph} \approx 1000$ K brown dwarfs out to 1 kpc and $T_{ph} \approx 500$ K brown dwarfs out to 100 pc, facilitating the science opportunities described above. Synoptic surveys will be important, as the proper motion information provided by multi-epoch astrometry enables the identification of thick disk/halo objects through reduced proper motion selection (e.g., Salim & Gould 2002, *ApJ* **575**, L83). Efficient characterization of large brown dwarf samples will also require simultaneous, wide-field, multi-object NIR spectroscopy. A 100-fold increase in sensitivity implies up to 10,000-fold higher surface densities; currently, UKIDSS finds 1 T dwarf every 16 square degrees down to J=19 (Pinfield et al. 2008, *MNRAS* **390**, 304). EUCLID or NIR equivalents to SWIFT, Gemini/WFMOS or LAMOST will be essential for compiling large spectroscopic samples for cold brown dwarf population studies.

**Determining the Physical Characteristics of Individual Cold Brown Dwarfs**

Broad variation in the observed spectra of brown dwarfs—and, in the future, exoplanets—arises from the interplay of atmospheric properties and processes with mass, age, metallicity, external heating source, and possibly rotation and magnetic field strength. By disentangling the contributions of these various physical properties, it is possible in principle to uniquely characterize brown dwarfs on an individual basis. It has in fact become routine practice to extract temperature, surface gravity, metallicity, condensate cloud and vertical mixing parameters for individual L and T dwarfs using semi-empirical analyses of optical and infrared spectra (Burgasser et al. 2006, *ApJ* **639**, 1095; Leggett et al. 2007, *ApJ* **667**, 537; Cushing et al. 2008, *ApJ* **678**, 1372). Extending these studies to larger samples would facilitate the use of brown dwarfs as distance, age and chemical standards for Galactic populations. Indeed, brown dwarfs are already used to age-date 80-200 Myr clusters (via the lithium depletion boundary technique; Bildsten et al. 1997, *ApJ* **482**, 442; Stauffer et al. 1998, *ApJ* **499**, L199) and individual stellar systems of interest (Burgasser 2007, *ApJ* **658**, 617; Liu et al. 2007, *ApJ* **660**, 1507).

To realize these scientific opportunities, brown dwarf evolutionary and atmospheric models must be critically and precisely tested. Empirical benchmarks are key to this objective. One class of benchmarks, resolved brown dwarf companions to well-characterized stars, are relatively rare (2-3%; Lafreniere et al. 2007, *ApJ* **670**, 1367) but used extensively for detailed tests of atmosphere models (e.g., Leggett et al. 2008, *ApJ* **682**, 1256) and calibrating brown dwarf spectral characterization techniques (Burgasser et al. 2006, *ApJ* **639**, 1095; Warren et al. 2007, *MNRAS* **381**, 1400). Large-volume samples constructed from deep imaging surveys would expand the number of companion brown dwarfs known, as well as uncover even rarer brown dwarf-white dwarf and brown dwarf-subgiant pairs that have more reliable stellar age diagnostics (Pinfield et al. 2006, *MNRAS* **368**, 1281). Low-temperature age and composition benchmarks can also be found in older—and generally distant—coeval clusters (e.g., M67: ~4 Gyr, 800 pc; NGC 2509: ~8 Gyr, 900 pc). Again, deep NIR/MIR imaging surveys will be essential for identifying and characterizing these systems. Finally, "mass standards", short-period brown dwarf binaries for which complete astrometric and/or radial velocity orbits can be measured, can also be used for precise tests of models (Liu et al. 2008, *ApJ* **689**, 436). Advances in ground-based adaptive optics (AO) facilities over the past decade have enabled the first orbit measurements for a few young brown dwarf pairs, already indicating problems with evolutionary models up to ages of ~700 Myr (Zapatero Osorio et al. 2004, *ApJ* **615**, 958; Dupuy et al. 2009, *ApJ* in press). To expand samples of resolved brown dwarf systems and identify older, lower-temperature mass standards, continued improvements in AO capabilities toward higher sensitivities and angular resolutions, coupled with spectroscopic modes for component radial velocity and atmospheric characterization, are desirable.

Along with these empirical tests, there is a need for improvements in fundamental quantities such as chemical abundances and molecular opacities to advance low-temperature atmospheric models. Currently, there are large uncertainties in the absolute abundances of C, N and O in our reference standard, the Sun, with contemporaneous studies showing up to 70% variations in adopted values (Lodders 2003, *ApJ* **591**, 1220; Asplund et al. 2005, *ASP* **336**, 25). As CNO-bearing molecules are a dominant source of opacity in low-temperature atmospheres, abundance uncertainties translate into systematic uncertainties in atmosphere models, equivalent to varying the metallicity of a source by ~30%. Models also suffer from incompleteness in warm opacity line lists for key molecules. Current line lists for $CH_4$ are incomplete at wavelengths below 1.6 μm; the $NH_3$ line list in this region is non-existent (Freedman et al. 2008, *ApJS*, **174**, 504). These molecules produce strong, broad absorption bands in the infrared that lead to substantial flux redistribution, propagating spectral modeling errors to other wavelengths. The calculation of the very large number of transitions for these molecules is a computer-intensive task, and complementary laboratory studies are essential but currently challenging. Both efforts require sustained support if accurate models of cold brown dwarf and planetary atmospheres are to be realized.

**Key Infrastructure Needs for Cold Brown Dwarf Science in 2010-2020**

We conclude by summarizing in the Table below some of the key science infrastructure needs of the cold brown dwarf community to adequately address the science opportunities and investigations described above. We also list near-term or proposed projects that could potentially address these needs in the coming decade.

A common theme for future investigations is the need for deep, wide-field, synoptic NIR/MIR surveys probing down to ~1-10 μJy sensitivities, facilitating searches for the coldest brown dwarfs near the Sun and large-volume samples of warmer brown dwarfs to study Galactic populations. The abundant, panchromatic data provided by these surveys would be most efficiently disseminated and mined by the astronomical community with continued support and development of a National Virtual Observatory/Virtual Astronomical Observatory (NVO/VAO) network. The public availability of survey data and data-mining tools would also serve to level the playing field for institutions without access to major research facilities; e.g., some minority-serving institutions and state college programs. A coupled requirement to deep imaging surveys is the availability of sensitive NIR and (more importantly) MIR spectroscopic instrumentation on either space-based (e.g., JWST) or large ground-based platforms. Capabilities for wide-field, multi-object spectroscopy in the NIR are particularly needed to investigate large samples. Theoretical work on cold brown dwarf atmospheres in the coming decade will hinge largely on support for advanced computational facilities and laboratory programs to model detailed atmospheric processes (including condensate formation) and determine molecular opacities.

Investments in these infrastructure elements will support both cold brown dwarf and exoplanet science into the coming decade.

| Science Opportunity | Science Infrastructure Needs | Proposed Projects Addressing Needs |
|---|---|---|
| Identification and characterization of the lowest-temperature brown dwarfs (~300 K) | ~1-100 µJy NIR/MIR imaging surveys; compatible NIR/MIR spectroscopic capabilities; support for astronomical information science | WISE, LSST, Pan-STARRS, JDEM, SASIR, NIRSS, JWST (spectroscopy), SPICA/BLISS, NVO/VAO |
| Census of ~1000 K brown dwarfs in thick disk, halo, and 2-8 Gyr Galactic cluster populations | ~1-100 µJy NIR/MIR imaging/astrometric (synoptic) surveys; survey-mode NIR/MIR spectroscopy | WISE (with 2$^{nd}$ epoch), LSST, Pan-STARRS, JDEM, SASIR, NIRSS, EUCLID, NVO/VAO |
| Identification of age/mass/composition benchmarks | Deep NIR/MIR imaging/astrometric (synoptic) surveys; high angular resolution imaging (binary orbits) | WISE, LSST, Pan-STARRS, JDEM, SASIR, NIRSS, JWST (imaging), NGAO, NVO/VAO |
| Improved models of condensate cloud formation and low temperature atmospheric spectra | Support for advanced computational modeling; laboratory and theoretical opacity studies | |